\documentclass[a4paper,11pt]{article}
\pdfoutput=1 

\usepackage{jheppub} 

\usepackage[T1]{fontenc} 
\usepackage{appendix}
\def\bs{\ensuremath\boldsymbol}
\DeclareMathOperator{\tr}{tr}
\newenvironment{aleq}
    {\begin{equation}\begin{aligned}}
    {\end{aligned}\end{equation}\ignorespacesafterend}
\setcitestyle{square}

\title{\boldmath A Note on 4d Kination and Higher-Dimensional Uplifts }

\author[a]{Fien Apers,}
\author[a]{Joseph P. Conlon,}
\author[a]{Martin Mosny}

\affiliation[a]{Rudolf Peierls Centre for Theoretical Physics,\\Beecroft Building,\\ Parks Rd, Oxford, OX1 3PU,\\UK}

\emailAdd{fien.apers@physics.ox.ac.uk}
\emailAdd{joseph.conlon@physics.ox.ac.uk}
\emailAdd{martin.mosny@physics.ox.ac.uk}

\abstract{This note expands on the details of the relationship between 4d kination solutions in string cosmology and Kasner solutions of 10d general relativity. It extends previous analyses of this relationship in IIB string theory to other string theories and also to 11d M-theory, while also providing extensive detail on the relationship between perturbations of the 10d Kasner metric and the presence of radiation and matter backgrounds in the dimensionally reduced 4d kination theory.}

\begin{document} 
\maketitle
\flushbottom

\section{Introduction}

An important task for string theory is to make connections to observational physics. One of the most promising places where such links may be made is in the context of cosmology. The normal approach to this starts through dimensional reduction of string theory into a 4-dimensional effective field theory; all actual calculations are then performed within this 4-dimensional theory and it is only rarely that the underlying 10-dimensional theory is considered explicitly. In this respect, a detailed recent review of string cosmology is \cite{Cicoli:2023opf} (see \cite{Baumann:2009ni, Brandenberger:2023ver} for other reviews with a slightly different focus).

One potential non-standard cosmological epoch in the early universe is that of a kination epoch (the name first appears in \cite{Joyce:1996cp}), an epoch dominated by the kinetic energy of a rolling scalar field. Such epochs naturally arise in string theory when moduli fields roll down steep potentials, for example when running from the centre of moduli space to the boundaries (a recent detailed analysis of this is \cite{Apers:2024ffe}; see \cite{Conlon:2022pnx, Revello:2023hro, Shiu:2023fhb, Shiu:2023nph, Seo:2024qzf} for some other recent studies of kination epochs in string theory).

One intriguing aspect of kination epochs in the context of IIB string compactifications, realised in \cite{Apers:2022cyl}, is that the epoch also admits a direct 10-dimensional description in terms of a higher-dimensional Kasner solution. This provides an interesting contrast to most work in string cosmology, where no such 10-dimensional description exists. This provides one of the few examples where it is possible to go beyond the 4d description and work directly within the 10d theory (another famous example where a 10d description exists is the GKP stage of IIB flux compactifications \cite{Giddings:2001yu}, prior to the step of K\"ahler moduli stabilisation for which \cite{Kachru:2003aw} and \cite{Balasubramanian:2005zx} are the best-known examples).

The purpose of this note is to build on the analysis of \cite{Apers:2022cyl} and explore this link further. There are three sections. In the first, we extend the results of \cite{Apers:2022cyl} to other string theories beyond the IIB orientifold examples considered there. In the second, we fill in more details on the perturbations of the 10d theory and their connection to moduli backgrounds in the 4d kination theory, going beyond the overall Kasner dynamics associated to the volume mode that was considered in \cite{Apers:2022cyl}. In the final one, we explore the consequence of a gas of winding modes during the kination epoch.

\section{Kination In General String Theories with Multiple Rolling Moduli}

We first consider kination epochs in more general string scenarios than the type IIB orientifold examples with just a single rolling volume modulus considered in \cite{Apers:2022cyl}.
As a single framework to capture multiple examples, we consider a 4d EFT which contains two moduli, a volume modulus and a dilaton, both with vanishing potential.\footnote{Even if the potential does not strictly vanish, if it is steep enough, fields enter a kination epoch as they roll down it and the potential can effectively be neglected.} The kinetic energy for these two moduli are determined by a K\"ahler potential which takes the form
\begin{aleq}
    K = -\log s^m u^n,
\end{aleq}
where $m$ and $n$ are integers which will vary for different string theories, $u$ is (the real part of) the volume modulus and $s$ is (the real part of) the 4d dilaton (which in general will differ from the 10d dilaton by factors of volume). The canonically normalised moduli are given by
\begin{aleq}
    S = \sqrt{\dfrac{m}{2}} \log s, \quad U=  \sqrt{\dfrac{n}{2}} \log u,
\end{aleq}
in terms of which the Friedman equation and the scalar equations of motion are
\begin{aleq}
    & 6 H^2 = 6 \left(\dfrac{\Dot{a}}{a}\right)^2 = \Dot{S}^2 + \Dot{U}^2,\\
    & \Ddot{S} + 3H \Dot{S} = 0,\\
    & \Ddot{U} + 3H \Dot{U} = 0.\\
\end{aleq}
These are solved by
\begin{aleq}
\label{dws}
    H(t_E) = \dfrac{1}{3t_E}, \quad S(t_E) = S_0 + \dfrac{\alpha}{\sqrt{2/m}} M_p \ln \left( \dfrac{t_E}{t_0} \right), \quad U(t_E) = U_0 + \frac{\beta}{\sqrt{2/n}} M_p \ln \left( \dfrac{t_E}{t_0} \right),
\end{aleq}
subject to the condition
\begin{aleq}\label{condition}
   m \alpha^2 +n \beta^2 = \dfrac{4}{3}.
\end{aleq}
We write $t_E$ for the time coordinate to be explicit that this refers to a 4d Einstein frame metric.
This 4-dimensional theory, in standard 4d Einstein frame, has a kination metric with $a(t_E) \propto t_E^{1/3}$,
\begin{aleq}
\label{ffx}
    ds_4^2 = -dt_E^2 + a^2(t_E) dx_i dx^i = -dt_E^2 + t_E^{2/3} dx_i dx^i. \quad i = {1, 2, 3}
\end{aleq}
Eqs. (\ref{dws}) to (\ref{ffx}) describe the kination field and metric evolution in 4d Einstein frame.

In string theory compactifications, this 4d Einstein frame metric is related to the (4d components of the) 10d string frame metric as
\begin{equation}
    g_{4d, Einstein} = e^{-2\phi} \mathcal{V} g_{4d,s}.
\end{equation}
Based on this, we expect that the 4d kination solution (in 4d Einstein frame) should correspond to an uplifted 10-dimensional string frame solution of the form,
\begin{aleq}
    ds_{10}^2 = \left[e^{2\phi(t_E)} \mathcal{V}(t_E)^{-1}\right]\left( -dt_E^2 + t_E^{2/3} dx_idx^i \right) +  \left[\mathcal{V}^{1/3}(t_E)\right]ds_6^2,
\end{aleq}
where $e^\phi$ is the 10d rolling dilaton and $\mathcal{V}$ is the internal volume. 

With an appropriate redefinition of the time coordinate, we will confirm that this is equivalent to a Mueller metric \cite{MUELLER199037}  which is a generalisation of the Kasner metric of the form
\begin{aleq}\label{mueller1}
    ds^2 = -dt^2 + \sum_{i} t^{2p_i} dx_i^2, \quad e^\phi = t^p
\end{aleq}
to include a rolling dilaton, satisfying the conditions
\begin{aleq}\label{mueller2}
    \sum_i p_i = 1 + 2p, \quad \sum_i p_i^2 = 1.
\end{aleq}

\subsection{Kination and its 10d uplift in type IIA string theory}

Here we make this connection explicit for type IIA string theory, where
the dilaton and volume moduli $s$ and $u$ are defined by (for example, see \cite{Grimm:2004ua})
\begin{aleq}
    s = e^{-\phi} \sqrt{\mathcal{V}}, \quad u = \mathcal{V}^{1/3},
\end{aleq}
with the Kahler metric given by
\begin{aleq}
    K = - \log s^4 u^3.
\end{aleq}
Here the dilaton itself directly defines the relationship between the 4d Einstein frame metric and the string frame metric. If $t_E$ denotes the 4d Einstein frame metric with the standard kination scale factor $a(t_E) \sim t_E^{1/3}$, then the corresponding expected uplift of the 4d kination metric to a 10d string frame metric is
\begin{aleq}
    ds_{10}^2 = [s(t_E)]^{-2} (-dt_E^2 + a^2(t_E) dx_n dx^n) + [u(t_E)] ds_6^2.
\end{aleq}
As $s(t_E) \sim t_E^{\alpha}$ and $u(t_E) \sim t_E^{\beta}$, this gives 
\begin{aleq}\label{iia}
    ds_{10}^2 &= -t_E^{-2\alpha}dt_E^2 + t_E^{\frac{2(1-3\alpha)}{3}} dx_n dx^n + t_E^{\beta} ds_6^2\\
    &= - d\tau^2 + \tau^{\frac{2(1-3\alpha)}{3(1-\alpha)}} dx_n dx^n + \tau^{\frac{\beta}{1-\alpha}}ds_6^2,
\end{aleq}
where $\tau = t_E^{1-\alpha}$.
The string coupling $e^{\phi}$ also depends on time as follows
\begin{aleq}
    e^{\phi(t_E)} = u(t_E)^{3/2}s(t_E)^{-1}= t_E^{\frac{1}{2}(3\beta - 2\alpha)} = \tau^{\frac{3\beta-2 \alpha}{2(1-\alpha)}}.
    \label{yup}
\end{aleq}
Eqs. (\ref{iia}) and (\ref{yup}) indeed represent a Mueller solution with
\begin{aleq}
    p_1 = \dfrac{1-3\alpha}{3(1-\alpha)}, \quad p_2 = \dfrac{\beta}{2(1-\alpha)}, \quad p = \dfrac{3\beta-2 \alpha}{2(1-\alpha)}.
\end{aleq}
These satisfy
\begin{aleq}\label{mueller}
    & 3p_1 + 6p_2-2p=1,\\
    & 3p_1^2 + 6p_2^2 = 1,
\end{aleq}
if and only if
\begin{aleq}
    4 \alpha^2 + 3 \beta^2 = \dfrac{4}{3},
\end{aleq}
which is precisely the condition \eqref{condition} applicable to the 4d kination solution. The $\alpha$ and $\beta$ parameters are arbitrary (subject to the condition of Eq. \eqref{condition}) and so can represent kinetic evolution in the direction of either or both of changing volume and changing dilaton.

\subsection{Kination and its 10d uplift for other string theories}

We can proceed similarly for other string theories, likewise obtaining 10d Mueller metrics for each case as the uplift of a kination solution involving rolling dilaton and volume moduli in the 4d effective field theory. The results are summarized in Table 1. 

\begin{table}[h!]
\begin{center}
    \setlength{\tabcolsep}{12pt} 
    \renewcommand{\arraystretch}{1.9} 
    \begin{tabular}{ | l | l | l | l |}
    \hline
    \textbf{Type IIA} & $m=4, n=3$ & $s= e^{-\phi} \mathcal{V}^{1/2}$& $u = \mathcal{V}^{1/3}$ \\
    \hline
     \multicolumn{1}{|c|}{} & \multicolumn{3}{l|}{$p_1 = \frac{1-3\alpha}{3(1-\alpha)}, \quad p_2 = \frac{\beta}{2(1-\alpha)}, \quad p = \frac{3\beta-2 \alpha}{2(1-\alpha)}$} \\
    \hline
    \textbf{Type IIB} & $m=1, n=3$ & $s= e^{-\phi} $ & $u = \mathcal{V}^{2/3}e^{-\phi}$ \\
     \hline
     \multicolumn{1}{|c|}{} & \multicolumn{3}{l|}{$ p_1 = \frac{3 \alpha +9 \beta -4}{3 (\alpha +3 \beta -4)}, \quad
p_2 =\frac{\alpha -\beta }{\alpha +3 \beta -4}, \quad
p = \frac{4 \alpha }{\alpha +3 \beta -4}$} \\
    \hline
    \textbf{Heterotic} & $m=1, n=3$ & $s= e^{-2\phi} \mathcal{V}$ & $u = \mathcal{V}^{1/3}$ \\
     \hline
     \multicolumn{1}{|c|}{} & \multicolumn{3}{l|}{$p_1 = \frac{2/3-\alpha}{2-\alpha}, \quad p_2 = \frac{\beta}{2-\alpha}, \quad p = \frac{3\beta-\alpha}{2-\alpha}$ } \\
    \hline
    \textbf{Type I} & $m=1, n=3$ & $s= e^{-\phi} \mathcal{V}$ & $u = e^{-\phi}\mathcal{V}^{1/3}$  \\
    \hline
     \multicolumn{1}{|c|}{} & \multicolumn{3}{l|}{$ p_1 = \frac{1}{3} \frac{4 - 3(3\alpha + \beta)}{4 - 3\alpha - \beta}, \quad
p_2 = \frac{2(\beta-\alpha)}{4 - 3\alpha - \beta}, \quad
p = \frac{2(\beta-3\alpha)}{4 - 3\alpha - \beta}$ } \\
     \hline
\end{tabular}
\caption{10d uplift of the 4d kination metric $ds_4^2 = -dt^2 + t^{\frac{2}{3}}dx_i dx^i$ in different string theories with rolling dilaton and volume moduli: type IIA, type IIB  heterotic, and type I string theory}
\end{center}
\end{table}

\subsection{The Mueller solution and the 11d M-theory Kasner solution}

Type IIA string theories have a direct uplift to M-theory, interpreting the string coupling as the size of the 11th dimensions (with strong string coupling corresponding to large radii). In this case, 
the 10d rolling dilaton Mueller solution of Eqs. \eqref{mueller1} and \eqref{mueller2}, uplifted to an 11-dimensional M-theory solution, becomes
\begin{aleq}
    ds_{11}^2 &= e^{-\frac{2\phi}{3}} \left[-dt^2 + \sum_{i=1}^{9} t^{2p_i} dx_i^2\right] +  e^{\frac{4\phi}{3}} dz^2\\
    &= -t^{-\frac{2p}{3}}dt^2 + \sum_{i=1}^{9} t^{2p_i-\frac{2}{3}p}dx_i^2 + t^{\frac{4}{3}p}dz^2\\
    &= -d\tau^2 +  \sum_{i=1}^{9} t^{\frac{2p_i-2p/3}{1-p/3}} dx_i^2 + t^{\frac{4p/3}{1-p/3}}dz^2\\
    &\equiv  -d\tau^2 +  \sum_{i=1}^{9} \tau^{2 \Tilde{p}_i} dx_i^2 + \tau^{2\Tilde{p}_{10}}dz^2, \\
 \end{aleq}
where $z$ is the coordinate on the M-theory circle, $\tau$ is defined as $\tau = t^{1-p/3}$ and numerical prefactors are dropped. 

This is indeed an 11d Kasner solution as from the first Mueller condition $ \sum_{i=1}^{9} p_i = 1 + 2p$, it follows that
\begin{aleq}
    \sum_{i=1}^{10} \Tilde{p}_i = 1.
\end{aleq}
Combining this with the second 10d condition $\sum_{i=1}^{9} p_i^2 = 1 $, we also get
\begin{aleq}
    \sum_{i=1}^{10} \Tilde{p}_i^2 = 1.
\end{aleq}

\section{Perturbations of 10d Kasner Solutions and their relation to 4d Moduli Backgrounds}

Given the interest of Kasner solutions as a 10d completion of kination solutions, it is worthwhile understanding their perturbations better (the paper \cite{Sugiyama:2020mbu} describes gravitational waves in the context of Kasner spacetimes including the Unruh effect). From a 4-d perspective, such perturbations correspond to the presence of additional matter/radiation content in the 4d theory.

In this section we give a detailed analysis of the relationship between perturbations of the 10d Kasner solution and the moduli fields that are present in the 4d kination theory (the time-dependent nature of the Kasner solution is what differentiates this from the familiar flat space treatment).

\subsection{10d Perturbations of the Kasner solution}

We start with an analysis of the 10d solution. Anticipating section \ref{sec32} where we relate perturbations of the Kasner metric to moduli backgrounds in the kination solution, it is useful to express 10d quantities in terms of the kination conformal time $\eta$.\footnote{As field perturbations in 4d theories are most naturally expressed in terms of the conformal time.} This relates to the time $t$ of the 10-dimensional Kasner solution through $\eta = \tfrac{3}{4} t^{4/3}$, in which case the Kasner metric takes the form
\begin{equation}
ds_{\text{Kasner}}^2 = \sqrt{\frac{3}{4\eta}}\bigg(-d\eta^2 + \sum_{i=1}^3 dx_i^2 \bigg) + \sqrt{\frac{4\eta}{3}}\sum_{\alpha=1}^6 dy_\alpha^2.
\end{equation}
Here $x_i$ are the large noncompact coordinates, while $y_j$ are the small compact coordinates. Working in the temporal gauge, where the off-diagonal temporal perturbations vanish, the massless perturbations of this 10-dimensional metric decouple at leading order into three types. From a lower-dimensional perspective, these correspond to complex structure perturbations, volume modulus perturbations, and radiation perturbations.

\subsubsection*{Complex structure perturbations}

We first consider perturbations in the torus complex structure, expanding in a small parameter $\lambda$, which in general take the form
\begin{equation}
ds^2 = ds^2_{\text{Kasner}} + \lambda f(\eta,\bs x) dy_\alpha^2 + \lambda g(\eta,\bs x) dy_{\alpha+1}^2 + 2\lambda h(\eta,\bs x)dy_{\alpha}dy_{\alpha+1}. 
\end{equation}
To leading order in $\lambda$, the Einstein equations for diagonal and off-diagonal contributions decouple. Additionally, the only non-trivial position dependent solution for the diagonal components must satisfy $g(\eta,\bs x) = -f(\eta,\bs x)$ (so that this is a complex structure perturbation and not a volume perturbation). The remaining equation of motion for both the diagonal and off-diagonal perturbations is
\begin{equation}
\nabla^2 f - \partial_\eta^2 f -\frac{1}{4\eta^2} f = 0,
\end{equation}
with the same equation applying for $h(\eta, \bs x)$. These have  exact solutions in terms of Bessel functions,
\begin{eqnarray}
f(\eta,\bs x) & = & e^{i\bs k\cdot \bs x}\sqrt \eta ( c_1 J_0(k\eta)+c_2 Y_0(k\eta)), \\
h(\eta,\bs x) & = & e^{i\bs k\cdot \bs x}\sqrt \eta ( d_1 J_0(k\eta)+d_2 Y_0(k\eta)).
\end{eqnarray}
where $k = |\bs k|$.

\subsubsection*{Volume modulus perturbations}

To find volume modulus perturbations, we consider an ansatz of the form
\begin{equation}
ds^2 = ds^2_{\text{Kasner}}+\frac{3\lambda}{4\eta}\sum_{i,j=1}^3f_{ij}(\eta,\bs x)dx_idx_j +\lambda f(\eta,\bs x) \sum_{\alpha=1}^6 m_\alpha dy^2_\alpha,
\end{equation}
with $\lambda$ again a small numerical parameter and where $m_\alpha$ are numerical coefficients, with
\begin{equation}
f_{ij}(\eta,\bs x) = \begin{pmatrix} f_{11} & f_{12} & f_{13} \\ f_{12} & f_{22} & f_{23} \\ f_{13} & f_{23} & f_{33}\end{pmatrix}(\eta, \bs x).
\end{equation}
This will be a volume perturbation of the compact tori if $\sum_\alpha m_\alpha \neq 0$. An analytic solution can be acquired by looking at late-time wavelike solutions, equivalent to large $k$ wave solutions, where the highest order derivative terms dominate. The nonvanishing components of Einstein's vacuum equations then reduce down to three sets
\begin{eqnarray}
\label{aaa}
\partial^2_\eta[(\Sigma_\alpha m_\alpha)  f + \sum_i f_{ii}] & = & 0, \\
\label{bbb}
\nabla^2 f_{ii} - \partial_\eta^2 f_{ii} + \partial_i^2\bigg[(\Sigma_\alpha m_\alpha)f+\sum_j f_{jj}\bigg]-2\sum_j \partial_i\partial_j f_{ij} & = & 0, \\
\label{ccc}
 \nabla^2 f - \partial_\eta^2 f & = & 0.
\end{eqnarray}
Equation \ref{ccc} fixes $f(\eta,\bs x)$ to the standard wave solution
\begin{equation}
f(\eta,\bs x) = A e^{ik\eta+i\bs k \cdot \bs x}.
\end{equation}
The remaining equations are solved by taking
\begin{equation}
f_{ij}(\eta,\bs x) = H_{ij}e^{ik\eta+i\bs k \cdot \bs x},
\end{equation}
where $H_{ij}$ is now an amplitude matrix. Eq. \ref{aaa} requires that $\tr H = -(\Sigma_\alpha m_\alpha)A$, while Eq. \ref{bbb} implies that $H_{ij}k_j = 0$. This shows that perturbations in the volume modulus are necessarily accompanied by perturbations in the noncompact part of the metric. During compactification and dimensional reduction, these noncompact perturbations correspond to the 4d dilaton, which is then re-absorbed into the final definition of the noncompact Einstein frame metric.

\subsubsection*{Radiation perturbations}

Finally, consider the cross-coupled off-diagonal perturbation
\begin{equation}
ds^2 = ds_{\text{Kasner}}^2 + \frac{2\lambda}{\sqrt \eta}\sum_i f_i(\eta,\boldsymbol  x) dx_i dy_\alpha
\end{equation}
for some $\alpha$. As per the original 5d Kaluza–Klein compactification, this would correspond to a $U(1)$ radiation mode in the 4d theory. The resulting equations to first order in $\lambda$ are given by
\begin{eqnarray}
\sum_i \partial_i[f_i - \eta \partial_\eta f_i] & = & 0, \\ 
\nabla^2 f_i - \partial_\eta^2 f_i - \sum_j \partial_i \partial_j f_j & = & 0.
\end{eqnarray}
This has an exact solution of the form
\begin{equation}
f_i(\eta,\boldsymbol x) = A_i e^{ik \eta+ik\cdot x}, \ \ \ \ \ \ \boldsymbol A\cdot \boldsymbol k = 0,
\end{equation}
 where the polarization condition is equivalent to $\Sigma_i \partial_i f_i=0$. 

As a check on the above three types of perturbation, we can count the number of degrees of freedom. In the temporal gauge there are a total of 45 free metric components. The equations of motion impose six polarization conditions on the radiation perturbations $\boldsymbol A_\alpha\cdot \boldsymbol k = 0$, a single trace condition on volume perturbations $\sum_\alpha m_\alpha = -\tr H$, and three polarization conditions on the noncompact gravitational waves $k_i H_{ij} = 0$. This leaves $45-6-3-1 = 35$ degrees of freedom, matching the number of degrees of freedom of a graviton in ten dimensions. 

\subsubsection*{Massive Kaluza–Klein perturbations}

We can also look at perturbations of the Kasner metric which would correspond to massive modes from a 4-dimensional perspective.
Such massive KK modes can be obtained using the metric 
\begin{equation}
ds^2 = ds_{\text{Kasner}}^2 + \frac{2\lambda}{\sqrt \eta} e^{2\pi in\cdot y} f(\eta,\bs x) \sum_{i=1}^3\sum_{\alpha=1}^6 m_{i\alpha} dx_i dy_\alpha,
\end{equation}
for some vector $n_\alpha$. This has wavelike solutions with the late-time behaviour
\begin{equation}
f(\eta,\bs x) = A e^{ik\eta+i\bs k \cdot \bs x},
\end{equation}
together with the conditions $\Sigma_i k_im_{i\alpha} = 0$ for all $\alpha$ and $\Sigma_\alpha m_{i\alpha}n_\alpha = 0$ for all $i$. 

\subsection{10d Perturbations as 4d Fields on the Kination Background}
\label{sec32}

When expanding around a 10-dimensional background metric $\bar g_{MN}$ that solves the Einstein equation as $g_{MN} = \bar g_{MN}+H_{MN}$, then up to cubic contributions $\mathcal O(H^3)$ the Einstein–Hilbert action can be rewritten as the Fierz–Pauli action in curved spacetime \cite{Fierz:1939,Ortin:2015} 
\begin{equation}
\begin{split}
S = -\frac{1}{2\kappa^2_{10}}\int d^{10}x \sqrt{-\bar g} \bigg[& \frac{1}{4}\bar \nabla_M H_{AB}\bar \nabla^M H^{AB}-\frac{1}{2}\bar \nabla_M H_{AB}\bar \nabla^A H^{MB} \\
& +\frac{1}{2}\bar \nabla_M H^{MN}\bar \nabla_N H -\frac{1}{4}\bar \nabla_M H \bar \nabla^M H \bigg],
\end{split}
\end{equation}
where $H = H_A{}^A$. In this case, the background metric is the Kasner metric $g_{MN}= (g_{\mu\nu}, g_{ij}) = (\zeta^{-1} \eta_{\mu\nu}, \zeta \delta_{ij})$, where $\zeta = \sqrt{4\eta/3}$. The bar over the covariant derivative indicates that this is the ten dimensional covariant derivative as opposed to the four dimensional covariant derivative that will appear after compactification.

It is convenient to rewrite the perturbations as 
\begin{equation}\label{metric_decomp}
H_{MN} \equiv
\begin{pmatrix}
h_{\mu\nu}-g_{\mu\nu}\phi & \zeta A_{\mu i} \\
\zeta A_{\nu j} & 2\zeta \phi_{ij}
\end{pmatrix}, \ \ \ \ \ \ \ \ \ \ H^{MN} \equiv 
\begin{pmatrix}
h^{\mu\nu}-g^{\mu\nu}\phi & \zeta A^{\mu i} \\
\zeta A^{\nu j} & 2\zeta \phi^{ij}
\end{pmatrix},
\end{equation}
where $\phi = \delta^{ij} \phi_{ij}$. The Kasner compactification differs from a standard compactification on a torus by the time-dependent factors of $\zeta$ \cite{Han:1998sg}. The expressions for the four terms in the Fierz–Pauli action in terms of the metric decomposition Eq. \eqref{metric_decomp} are found in the appendix.

To extract an action for the massless perturbations, we assume that the perturbations are independent of the compact coordinates. In this case, we can directly integrate over the compact manifold to acquire a volume term $\mathcal V_6 =\mathcal V_{6,0} \zeta^3$. Meanwhile, expanding the Fierz–Pauli action yields what at first appears to be a mass term for the gauge field. However, this ends up forming part of a total derivative,
\begin{eqnarray}
\frac{2\kappa_{10}^2}{\mathcal V_{6,0}\sqrt{-g}}\mathcal L & \supset & \frac{2}{9}\zeta^2 A^2 + \frac{2}{3}\zeta^4 A^{\nu i}\nabla_0 A_{\nu i} \\
& = & \frac{2}{9}\zeta^2 A^2 +\frac{2}{3}\bigg(\frac{1}{2}\nabla_0(\zeta^4 A^{\nu i}A_{\nu i}) - \zeta^2 A^2\bigg) \\
& = &  \frac{2}{9}\zeta^2 A^2 + \frac{2}{3}\bigg(\frac{1}{2}\nabla_\mu X^\mu -\frac{1}{3}\zeta^2 A^2\bigg) \\
& = & \frac{1}{3}\nabla_\mu X^\mu,
\end{eqnarray}
where $A^2 = A^{\nu i}A_{\nu i}$ and $X^\mu = \delta^\mu_0 \zeta^4 A^2$. Dropping this total derivative leaves, as expected, a massless gauge field. A similar argument shows that the terms $\phi \nabla_0 \phi$ and $\phi^{ij}\nabla_0 \phi_{ij}$ also drop out the action as total derivatives.

The resulting Lagrangian is given by
\begin{equation}
\label{drty}
\begin{split}
\mathcal L = \frac{\sqrt{-g}}{2\kappa_{4}^2}\zeta^3\bigg[& \tfrac{1}{4}\partial_\mu h \partial^\mu h -\tfrac{1}{2}\nabla_\mu h^{\mu \alpha} \nabla_\alpha h + \tfrac{1}{2}\nabla_\mu h_{\nu \rho}\nabla^\nu h^{\mu \rho} - 
 \tfrac{1}{4}\nabla_\mu h_{\nu \rho}\nabla^\mu h^{\nu \rho} \\
& - \tfrac{1}{2}\nabla_\mu \phi \nabla^\mu \phi - \tfrac{1}{4}\zeta^2 g^{ij}F_{i,\mu \nu}F^{\mu \nu}_j - \zeta^2 g^{ik}g^{jp}\nabla_\mu \phi_{ij}\nabla^\mu \phi_{kp} \\
& -\tfrac{4}{3}\zeta^{-1}\phi \nabla_0 h+\zeta^{-1}h_0^\mu \nabla_\mu h - \tfrac{8}{3}\zeta^{-1}h_0^\mu \nabla_\mu \phi\bigg],
\end{split}
\end{equation}
where $h = h^\mu{}_\mu$ and $\kappa_{4}^{-2} = \mathcal V_{6,0}\kappa_{10}^{-2}$ is the 4d gravitational coupling constant. Meanwhile, $F_{i,\mu\nu}$ are the field-strength tensors for the abelian $A_{i,\mu}$ gauge fields. To go into the Einstein frame requires performing a Weyl transformation on the metric $g_{\mu \nu} \rightarrow \mathcal \zeta^{-3}g_{\mu\nu}$ and its perturbations $h_{\mu\nu}\rightarrow \zeta^{-3}h_{\mu\nu}$. The Weyl transformation rescales each term in the second and third lines individually to eliminate the $\zeta^3$ prefactor. The first line transforms to cancel this prefactor while also giving rise to two total derivatives that can be dropped and a term that cancels with the $h_0^\mu\nabla_\mu h$ term on the third line. The resulting action is
\begin{equation}
\begin{split}
S = \frac{1}{2\kappa^2_{4}}\int d^4 x \sqrt{-g}& \bigg[\tfrac{1}{4}\partial_\mu h \partial^\mu h -\tfrac{1}{2}\nabla_\mu h^{\mu \alpha} \nabla_\alpha h + \tfrac{1}{2}\nabla_\mu h_{\nu \rho}\nabla^\nu h^{\mu \rho} - 
 \tfrac{1}{4}\nabla_\mu h_{\nu \rho}\nabla^\mu h^{\nu \rho}\\
 &-\tfrac{1}{2}\partial_\mu \phi \partial^\mu \phi - \tfrac{1}{4\tilde g^2}\sum_{i=1}^6 F_i^2 - \sum_{ij}\partial_\mu \phi_{ij}\partial^\mu \phi_{ij} \\
& + \tfrac{1}{\sqrt 3}\phi B^\mu \partial_\mu h + \tfrac{2}{\sqrt 3}B^\mu h^\nu_\mu \partial_\nu \phi\bigg],
\end{split}
\end{equation}
where we introduce the vector field $B^\mu = -\tfrac{4}{\sqrt 3}(\zeta^{-4},0,0,0)$ and $\tilde g^2 = \zeta^{-1}\mathcal \zeta^{-3} \sim \eta^{-2}$ is a time-dependent coupling for the gauge field.

In order to simplify the equations, we next separate out the 4d dilaton from the complex structure scalars $\phi_{ij}$, which can be done via a field redefinition
\begin{equation}
\begin{split}
\varphi & = \tfrac{2}{\sqrt 3}\phi, \\
\Delta_1 & = \phi_{11}-\phi_{22}, \\
\Delta_2 & = \phi_{33}-\phi_{44}, \\
\Delta_3 & = \phi_{55}-\phi_{66}, \\
\Delta_4 & = \tfrac{1}{\sqrt 2}(\phi_{33}+\phi_{44}-\phi_{11}-\phi_{22}), \\
\Delta_5 & = \tfrac{1}{\sqrt  6}(2\phi_{33}+2\phi_{44}-\phi_{11}-\phi_{22}-\phi_{55}-\phi_{66}).
\end{split}
\end{equation}
The final form of the 4d Einstein frame action is then given by
\begin{equation}\label{mainaction}
\begin{split}
S = \frac{1}{2\kappa^2_4}\int d^4 x \sqrt{-g}& \bigg[\tfrac{1}{4}\partial_\mu h \partial^\mu h -\tfrac{1}{2}\nabla_\mu h^{\mu \alpha} \nabla_\alpha h + \tfrac{1}{2}\nabla_\mu h_{\nu \rho}\nabla^\nu h^{\mu \rho} - 
 \tfrac{1}{4}\nabla_\mu h_{\nu \rho}\nabla^\mu h^{\nu \rho}\\
 &-\tfrac{1}{2}\partial_\mu \varphi \partial^\mu \varphi - \tfrac{1}{4\tilde g^2}\sum_{i=1}^6 F_i^2 - \tfrac{1}{2} \sum_{i=1}^5 \partial_\mu \Delta_i \partial^\mu \Delta_i - \sum_{i\neq j}\partial_\mu \phi_{ij}\partial^\mu \phi_{ij} \\
& + \tfrac{1}{2}\varphi B^\mu \partial_\mu h + B^\mu h^\nu_\mu \partial_\nu \varphi\bigg].
\end{split}
\end{equation}
The components of the action involving the metric perturbations $h_{\mu\nu}$ and the dilaton $\varphi$ match exactly the action found by expanding out the kination action with $\Phi = \bar \Phi+\varphi$ and $g_{\mu\nu} \rightarrow g_{\mu\nu}+h_{\mu\nu}$. This calculation is given in the appendix.

\subsubsection*{Gauge transformations in 10d and 4d}

The ten dimensional gauge transformation of the Fierz–Pauli action are given by
\begin{equation}
H_{MN}\rightarrow H_{MN}+\bar \nabla_M \xi_N+\bar \nabla_N \xi_M.
\end{equation}
These can be decomposed into gauge transformations on the various fields, where the Weyl transformation also transforms the gauge parameter as $\xi_\mu \rightarrow \zeta^{-3}\xi_\mu$. Along with the $\xi_i \rightarrow \zeta \xi_i$, this gives the 4d gauge transformations as
\begin{equation}
\begin{split}
h_{\mu \nu}& \rightarrow h_{\mu\nu}+\nabla_\mu \xi_\nu + \nabla_\nu \xi_\mu, \\
A_{\mu i} & \rightarrow A_{\mu i}+\partial_\mu \xi_i, \\
\varphi & \rightarrow \varphi - \tfrac{4}{\sqrt 3}\zeta^{-4}\xi_0.
\end{split}
\end{equation}

The $\xi_i$ components result in gauge transformations for the gauge field and leave the $F_i^2$ term invariant. The gauge parameter $\xi_\mu$ corresponds to the active coordinate transformations $x^\mu \rightarrow x^\mu + \xi^\mu$, which act on the metric in the familiar way. They also act on the dilaton as expected in a kinating background. Under the active coordinate transformation the kination field can be related as $\Phi(x^\mu) = \Phi'(x^\mu - \xi^\mu)$. If the dilaton is a small perturbation around a background kinating field $\Phi = \bar \Phi+ \varphi$ then
\begin{equation}
\bar \Phi(\eta) + \varphi(x^\mu) = \bar \Phi(\eta - \xi^0)+\tilde\varphi(x^\mu - \xi^\mu).
\end{equation}
Taylor expanding the background $\bar \Phi(\eta-\xi^0)$, implies that the dilaton transformations as
\begin{equation}
\varphi(x^\mu) \rightarrow \varphi'(x^\mu) = \varphi(x^\mu) - \zeta^{-2}\bar \Phi'\xi_0.
\end{equation}
Since $\bar \Phi'(\eta) = \tfrac{4}{\sqrt 3}\zeta^{-2}$, this yields the same transformation as acquired from the compactification of the Kasner solution. In the appendix, we illustrate the explicit gauge invariance of 4d action for one of these transformations.

\subsubsection*{Equations of motion}

In conformal time, the 4d kination metric takes the form
\begin{equation}
ds^2_{\text{Kination}} = \frac{4\eta}{3}(-d\eta^2+dx_i^2).
\end{equation}
In the late-time limit where the mixing between the 4d dilaton and the graviton, the last line of \eqref{mainaction}, become negligible, all massless scalar fields satisfy the equation of motion
\begin{equation}
\frac{1}{\sqrt{-g}}\partial_\mu (\sqrt{-g}g^{\mu\nu}\partial_\nu \phi) = 0,
\end{equation}
which gives the equation
\begin{equation}\label{scalareq}
\partial^2_\eta \phi + \eta^{-1} \partial_\eta \phi = \nabla^2 \phi.
\end{equation}
This equation can be solved in terms of Bessel functions to give
\begin{equation}
\phi(\eta, \bs x) = e^{i\bs k \cdot \bs x}(c_1 J_0(k\eta)+c_2 Y_0(k\eta)).
\end{equation}

For the massless gauge field arising from dimensional reduction of the mixed off-diagonal compact-noncompact part of the metric, its equation of motion in the temporal gauge is given by
\begin{equation}
\tfrac{2}{\eta}\partial_0 A_i + \partial_0^2 A_i -\nabla^2 A_i = 0,
\end{equation}
which has a solution of the form
\begin{equation}
A_i(\eta, \bs x) = \tfrac{C_i}{\eta}e^{ik\eta+i\bs k \cdot \bs x}
\end{equation}
together with $\bs C\cdot \bs k = 0$. The $\eta^{-1}$ prefactor arises due to the time-dependent coupling constant $\tilde g$. It can be eliminated through a field redefinition at the expense of introducing terms in the action that are negligible in the late time/small distance limit.

\subsubsection*{Energy density}

The energy density of the massless KK gauge fields for a normal observer $n_\mu = (-\zeta,0,0,0)$ can be calculated from the action by deriving the stress-energy tensor
\begin{equation}
\rho = n_\mu n_\nu T^{\mu\nu} \sim - \frac{\zeta^2}{\tilde g^2}(F^{0\mu}F^0{}_\mu - \tfrac{1}{4}g^{00}F^2).
\end{equation}
Plugging in our solution to the gauge field and averaging over the oscillations yields a energy density that behaves as
\begin{equation}
\rho \propto c_1 k^2 a^{-4}+ c_2 a^{-8}.
\end{equation}
This shows that the energy density indeed behaves as radiation for high energy modes $k\gg 1$ (which are well within the horizon and admit an ordinary particle interpretation) but behaves as $\rho \propto a^{-8}$ for low energy modes (outside the horizon). A similar calculation for the scalar field perturbations shows that at late times these also behave as radiation $\rho \propto k^2a^{-4}$.

In the string frame the mass of heavy KK modes goes as $m_{KK} \propto \zeta^{-1/2}$ which results in a mass in the Einstein frame of
\begin{equation}
m_{KK} \propto \mathcal V_6^{-2/3} \propto a^{-2}.
\end{equation}
The winding modes in the meantime have a time dependence of the form $m_{w} \propto \zeta^{1/2}$, which in the Einstein frame is
\begin{equation}
m_w \propto \mathcal V_6^{-1/3} \propto a^{-1}.
\end{equation}
The energy density goes as $\rho = nm$ where $n$ is the number density that scales as the inverse volume, giving
\begin{equation}
\rho_{KK}\propto a^{-5}, \ \ \ \ \ \ \rho_w \propto a^{-4}.
\end{equation}
We thus see that winding mode states decay as radiation while KK modes decay even faster (a consequence of the fact that the mass of these particles also evolves with time and so these heavy modes cannot be treated as simply dust).

\section{Winding Mode Domination}

Modes with non-trivial winding in the extra dimensions are an interesting example of states whose mass grows relative to the string scale as the size of the extra dimensions increases (note that although non-toroidal Calabi–Yau compactifications do not allow for winding modes with integer quantum numbers, discrete $\mathbb{Z}_2$ winding modes are still possible). We explore here the evolution of the 10-dimensional Kasner solution in the presence of a gas of these. We assume that the compact dimensions contain an averaged gas of such modes, such that we can roughly deal with the effect on the winding strings semi-classically. We also assume, for simplicity, a toroidal geometry for the extra dimensions.

Classical cosmic strings with tension $\mu$ pointing in the $y_6$ direction have a stress energy tensor of
\begin{equation}
T^\mu{}_\nu = \mu \, \text{diag}[1,0,0,0,\dots, -1]\delta(x_1)\delta(x_2)\cdots \delta (y_5).
\end{equation}
If we have a large number of such winding modes with a number density $n$ along each direction that we can average over, and assuming a metric of the form
\begin{equation}
ds^2_{10} = -dt^2+p(t)d\bs x^2 +q(t)d \bs y^2,
\end{equation}
then the indices-lowered stress energy tensor is
\begin{equation}
T_{\mu\nu} = \mu n \ \text{diag}[6,0,0,0,-q,-q,-q,-q,-q,-q].
\end{equation}
Writing $p(t) = e^{\alpha(t)}$ and $q(t) = e^{\beta(t)}$, the 10d Einstein equations reduce to
\begin{gather}
\dot \alpha^2 + 6\dot \alpha \dot \beta + 5 \dot \beta^2 = 64 \pi \mu n_0 e^{-\tfrac{3}{2}\alpha-\tfrac{5}{2}\beta}, \\
3\dot \alpha^2 +12 \dot \alpha \dot \beta + 21 \dot \beta^2 + 4 \ddot \alpha + 12 \ddot \beta = 0, \\
6 \dot \alpha^2 + 15 \dot \alpha \dot \beta + 15 \dot \beta^2 + 6 \ddot \alpha + 10 \ddot \beta = 32 \pi \mu n_0 e^{-\tfrac{3}{2}\alpha-\tfrac{5}{2}\beta}.
\end{gather}
These allow us to solve for $\alpha$ and $\beta$ along with fixing $n = n_0 /\sqrt{p^3q^5}$, which is the scaling relation required for winding modes whose number density does not change in the direction that they are winding. 

These equations can be solved for in the early and late-time limits. Early times with $n_0 \sim 0$ give a 10 Kasner / 4d kination solution. Numerically one can see that at late times $\dot \alpha = 3\dot \beta$, which holds for any $n_0$. This then implies that $\ddot \beta = -7\dot \beta^2/2$ and $\ddot \alpha= -7\dot \alpha^2/6$. With this the equations simplify to
\begin{equation}
\ddot \beta = 2\pi \mu n,
\end{equation}
with $\beta = \alpha/3-c/3$, where $c$ is an integration constant. We can then solve these equations to get
\begin{gather}
q(t) = \bigg[\frac{7}{2}\sqrt{2\pi \mu n_0 e^{-3c}}t + \tilde c\bigg]^{2/7}, \\
p(t) = e^c \bigg[\frac{7}{2}\sqrt{2\pi \mu n_0 e^{-3c}}t + \tilde c\bigg]^{6/7}.
\end{gather}
Roughly, the late-time behaviour here is $q(t) \sim t^{2/7}$ and $p(t) \sim t^{6/7}$.

Compactification of the metric gives an Einstein frame metric of
\begin{equation}
ds_E^2 = -q^3 dt^2 + q^3 p d\bs x^2,
\end{equation}
hence $a(t) = \sqrt{q^3 p} \sim t^{6/7}$. To get the metric into the FRW form we need to perform a change of temporal coordinates $\tau = f(t)$ with $q^3 = \dot f^2$. Then the Hubble constant is defined by
\begin{equation}
H(\tau) = \frac{1}{a}\frac{d}{d\tau}a(\tau) = \frac{\dot a}{a}\frac{d\tau}{dt} = \frac{\dot a}{a\dot f} = \frac{\dot a}{a q^{3/2}}.
\end{equation}
The energy density during this winding mode domination behaves as $\rho \propto H^2$, giving
\begin{equation}
\rho \propto a^{-10/3}.
\end{equation}
Notice that this scales slower than radiation but faster than matter. The same result can also be understood more directly by noting that the energy density of winding modes behaves as $\rho = m_w n_3 \propto m_w/a^3$. Compactification of the action shows that the 10d mass $m_{10} = \mathcal V^{1/6}\mu$ gets rescaled due to the Weyl transformation to $m_w = \mu \mathcal V^{-1/3} \propto q^{-1}$. Since at late times $q \propto a^{1/3}$, this gives the same energy density behaviour. 

From the Friedmann equation one can see that the resulting winding mode domination cosmology is equivalent to one with a cosmological fluid with $w = 1/9$ (although we would not expect such modes to remain stable; they should ultimately decay to lighter modes).

\section{Conclusions}

This note has extended the previous work of \cite{Apers:2022cyl} on the relationship between kination and Kasner solutions in type IIB compactifications. Here we have provided much more detail on the kination-Kasner relationship. In the first part of this paper, we extended the earlier analysis of \cite{Apers:2022cyl} to multiple moduli and other corners of the M-theory duality web, such as type I, heterotic, IIA and also M-theory. 

As $\rho_{kin} \sim a^{-6}$ whereas $\rho_{\gamma} \sim a^{-4}$, kination backgrounds are unstable against a small initial radiation fraction. Moduli excitations in the 4d theory correspond to perturbations of the 10d metric. To understand these better,
in the second part of the paper we went into very explicit detail on the correspondence between perturbations of the 10d Kasner solution and the moduli fields in the 4d kination solution.

Post-inflationary kination epochs are well motivated in string theory and offer an appealing way to modify the standard cosmology. By providing more explicit detail on them, and their relationship to higher-dimensional Kasner solutions, in this note we hope to have provided foundations that can subsequently be used to improve our understanding of such epochs, with the ultimate goal of extracting testable predictions.

\acknowledgments

We thank Ethan Carragher, Ed Copeland, Filippo Revello and Gonzalo Villa for discussions related to this topic. FA is supported by the Clarendon Scholarship in partnership with the Scatcherd European Scholarship, Saven European Scholarship, and the Hertford College Peter Howard Scholarship. JC acknowledges support from the STFC consolidated grants ST/T000864/1 and ST/X000761/1 and is also a member of the COST Action COSMIC WISPers CA21106, supported by COST (European Cooperation in Science and Technology). JC thanks the Erwin-Schr\"odinger-Institut in Vienna for hospitality during the writing up of this work. MM is supported by the St John’s College Graduate Scholarship in partnership with STFC. For the purpose of Open Access, the authors have applied a CC BY public copyright licence to any Author Accepted Manuscript version arising from this submission.

\appendix

\section{Appendix}

Here we provide various formulae that may be useful as reference but would obstruct the flow of the main text.

\subsection*{Formulae for Covariant Derivatives of the Metric}

Here we include various formulae for the covariant derivatives of the metric that are useful for summing the terms in the Fierz–Pauli action and deriving Eq. \ref{drty}.
\begin{equation}
\begin{split}
& \bar \nabla_\mu H_{\nu \rho} = \nabla_\mu h_{\nu \rho} - g_{\nu \rho}\nabla_\mu \phi, \\
& \bar \nabla_\mu H_{\nu i} = \zeta\nabla_\mu A_{\nu i} -\tfrac{1}{3} g_{\mu 0} A_{\nu i}\\
& \bar \nabla_\mu H_{ij} = 2\zeta \nabla_\mu \phi_{ij}, \\
& \bar \nabla_i H_{\mu \nu} = \partial_i h_{\mu \nu} - g_{\mu\nu}\partial_i \phi + \tfrac{1}{3}g_{\mu 0}A_{\nu i} +\tfrac{1}{3}g_{\nu 0}A_{\mu i}, \\
& \bar \nabla_i H_{\mu j} = \zeta \partial_i A_{\mu j} + \tfrac{1}{3}g_{\mu 0}[2\phi_{ij}+\zeta^{-1} g_{ij}\phi] - \tfrac{1}{3}\zeta^{-1}g_{ij}h_{0\mu}, \\
& \bar \nabla_i H_{jk} = 2\zeta \partial_i \phi_{jk} - \tfrac{1}{3}g_{ij}  A_{0k} - \tfrac{1}{3}g_{ik} A_{0j}.
\end{split}
\end{equation}
and 
\begin{equation}
\begin{split}
& \bar \nabla^\mu H^{\nu \rho} = \nabla^\mu h^{\nu \rho} - g^{\nu \rho}\nabla^\mu \phi, \\
& \bar \nabla^\mu H^{\nu i} = \zeta g^{ij}\nabla^\mu A^\nu{}_j - \tfrac{1}{3}\delta^\mu_0 A^{\nu i}\\
& \bar \nabla^\mu H^{ij} = 2 \zeta g^{ik}g^{jm}\nabla^\mu \phi_{km}, \\
& \bar \nabla^i H^{\mu \nu} = \partial^i h^{\mu \nu} - g^{\mu\nu}\partial^i \phi + \tfrac{1}{3}\delta^\mu_0A^{\nu i} + \tfrac{1}{3}\delta^\nu_0 A^{\mu i}, \\
& \bar \nabla^i H^{\mu j} = \zeta \partial^i A^{\mu j} + \tfrac{1}{3}\delta^\mu_0[2\phi^{ij} + \zeta^{-1}g^{ij}\phi] - \tfrac{1}{3}\zeta^{-1}g^{ij}h_0^\mu, \\
& \bar \nabla^i H^{jk} = 2\zeta \partial^i\phi^{jk} - \tfrac{1}{3}g^{ij}A_0{}^k - \tfrac{1}{3}g^{ik}A_0{}^j,
\end{split}
\end{equation}
where Latin indices were raised with $g^{ij}$ while Greek ones with $g^{\mu\nu}$. We have to be careful not to commute $g^{ij}$ past $\nabla_\mu$ covariant derivatives since only the non-compact metric commutes with these.

Next, $H = h - 2 \phi$, where $h = \zeta\eta^{\mu\nu}h_{\mu\nu} = g^{\mu \nu}h_{\mu\nu}$, and so
\begin{equation}
-\tfrac{1}{4}\bar \nabla_M H \bar \nabla^M H = -\tfrac{1}{4}\partial_\mu h \partial^\mu h + \partial_\mu h \partial^\mu \phi - \partial_\mu \phi \partial^\mu \phi -\tfrac{1}{4} \partial_i h \partial^i h +\partial_i h \partial^i \phi - \partial_i \phi \partial^i \phi.
\end{equation}
It is also useful to evaluate, up to integration by parts, that
\begin{equation}
\begin{split}
\tfrac{1}{2}\bar \nabla_M H^{MN} \bar \nabla_N H & = \tfrac{1}{2}\nabla_\mu h^{\mu \nu} \nabla_\nu h - \nabla_\mu h^{\mu \nu}\nabla_\nu \phi - \tfrac{1}{2}\nabla_\mu h \nabla^\mu \phi +  \nabla_\mu \phi \nabla^\mu \phi  \\
& + \tfrac{4}{3}\zeta^{-1}\phi \nabla_0 h - \tfrac{8}{3} \zeta^{-1}\phi \nabla_0 \phi - \zeta^{-1} h_0^\mu \nabla_\mu h + 2 \zeta^{-1}h_0^\mu \nabla_\mu \phi\\
& - \tfrac{4}{3}A_0^i \partial_i h + \tfrac{8}{3} A_0^i \partial_i \phi \\
& +\tfrac{1}{2} \zeta \nabla_\mu A^\mu_i \partial^i h - \zeta \nabla_\mu A^\mu_i \partial^i \phi + \tfrac{1}{2}\zeta \partial_i A^{\mu i}\nabla_\mu h - \zeta \partial_i A^{\mu i}\nabla_\mu \phi \\
& - \zeta \partial_i \phi^{ij}\partial_j h + 2\zeta \partial_i \phi^{ij}\partial_j \phi,
\end{split}
\end{equation}
and also
\begin{equation}
\begin{split}
-\tfrac{1}{2}\bar \nabla_M H_{AB} \bar \nabla^A H^{MB} & = -\tfrac{1}{2} \nabla_\mu h_{\nu \rho}\nabla^\nu h^{\mu \rho} +\nabla_\mu h^{\mu \nu}\nabla_\nu \phi -\tfrac{1}{2} \nabla_\mu \phi \nabla^\mu \phi -\tfrac{1}{2} \zeta^2 g^{ij}\nabla_\mu A_{\nu i}\nabla^\nu A^\mu_j \\
& - \tfrac{1}{9}\zeta^{-1}A_{\mu i}A^{\mu i} - \tfrac{4}{9}A_{0i}A_0{}^i +\tfrac{1}{9}\zeta^{-1}[5\zeta^{-2}\phi^2 + 2 \phi_{ij}\phi^{ij}] + \tfrac{8}{9}\zeta^{-2}h_{00}\phi- \tfrac{1}{3}\zeta^{-2}h_{0\mu}h_0{}^\mu \\
& -\tfrac{1}{3}\zeta A^{\mu i}\nabla_0 A_{\mu i} - \tfrac{4}{3} \zeta \phi^{ij}\nabla_0 \phi_{ij} - \tfrac{2}{3} g^{ij} \phi \nabla_0 \phi_{ij} + \tfrac{2}{3}h_0^\mu g^{ij}\nabla_\mu \phi_{ij} \\
& - \tfrac{2}{3}\partial^i \phi A_{0i} + \tfrac{4}{3}\zeta \partial_i \phi^{ij}A_{0j} \\
& - 2\zeta^2 \partial_i \phi_{jk}\partial^j \phi^{ik} -\tfrac{1}{2} \zeta^2 \partial_i A_{\mu j}\partial^j A^{\mu i} \\
& - \zeta \nabla_\mu A_{\nu i}\partial^i h^{\mu \nu} + \zeta \partial^i \phi \nabla_\mu A^\mu_i - 2\zeta^2 \nabla_\mu \phi_{ij}\partial^i A^{\mu j},
\end{split}
\end{equation}
and finally,
\begin{equation}
\begin{split}
\tfrac{1}{4}\bar \nabla_M H_{AB} \bar \nabla^M H^{AB} & = \tfrac{1}{4}\nabla_\mu h_{\nu \rho}\nabla^\mu h^{\nu \rho} - \tfrac{1}{2}\nabla_\mu h \nabla^\mu \phi + \nabla_\mu \phi \nabla^\mu \phi \\
& +\tfrac{1}{2}\zeta^2 g^{ij} \nabla_\mu A_{\nu i}\nabla^\mu A^\nu_j + \zeta^2 g^{ik}g^{jp} \nabla_\mu \phi_{ij}\nabla^\mu \phi_{kp} \\
& - \tfrac{1}{9}\zeta^{-1}A_{\mu i}A^{\mu i} + \tfrac{4}{9}A_{0i}A_0{}^i - \tfrac{1}{9}\zeta^{-1}[5\zeta^{-2} \phi^2+2\phi_{ij}\phi^{ij}] - \tfrac{8}{9}\zeta^{-2}h_{00}\phi + \tfrac{1}{3}\zeta^{-2}h_{0\mu}h^\mu_0 \\ 
& - \tfrac{1}{3}\zeta A^{\mu i}\nabla_0 A_{\mu i} - \tfrac{2}{3}A_{0i}\partial^i \phi + \tfrac{4}{3}\zeta \phi^{ij}\partial_i A_{0j} - \tfrac{2}{3}h_0^\mu\partial^i A_{\mu i} \\
& + \tfrac{1}{4}\partial_i h_{\mu \nu}\partial^i h^{\mu \nu} -\tfrac{1}{2}\partial_i h \partial^i \phi + \partial_i \phi\partial^i \phi + \tfrac{1}{2} \zeta^2 \partial_i A_{\mu j}\partial^i A^{\mu j} + \zeta^2 \partial_i \phi_{jk}\partial^i \phi^{jk}.
\end{split}
\end{equation}
Here we can freely integrate the compact coordinate by parts since the compact metric $\zeta \delta_{ij}$ is flat, and so all of its Christoffel symbols vanish, and so $\nabla_i = \partial_i$. 

\subsection*{Expansion of the Kination Action}

The kination action can be written as
\begin{equation}\label{kination_action}
S = \frac{1}{2\kappa_4^2} \int d^4x \sqrt{-g'}\bigg[R' - \frac{1}{2}g'^{\mu\nu}\partial_\mu \Phi \partial_\nu \Phi\bigg],
\end{equation}
where the scalar field can be canonically normalized through $\Phi \rightarrow \sqrt{2}M_P^{-1}\Phi$. To match its perturbations to those acquired from the compactification of the Kasner solution in Eq. \eqref{mainaction}, we need to expand this solution to second order in the perturbations $g_{\mu\nu}'= g_{\mu\nu}+h_{\mu\nu}$ and $\Phi=\bar \Phi +\varphi$. Here the kination metric is given by
\begin{equation}
g_{\mu\nu} = \tfrac{4\eta}{3}\eta_{\mu\nu},
\end{equation}
for which 
\begin{eqnarray}
R_{\mu\nu} = \tfrac{3}{2\eta^2}\delta_{\mu0}\delta_{\nu0}, \ \ \ \ \ \ R = -\tfrac{9}{8\eta^3}.
\end{eqnarray}
Meanwhile, 
\begin{equation}
\bar \Phi = \bar \Phi_0 + \sqrt 3 \ln(\eta/\eta_0).
\end{equation}
It is useful to introduce the vector field $B_\mu = \partial_\mu \bar \Phi = \tfrac{\sqrt{3}}{\eta}\delta_{\mu 0}$, which satisfies $R_{\mu\nu} = B_\mu B_\nu/2$ and $R = B^\mu B_\mu/2$.

To second order in metric perturbations the determinant is expanded out as \cite{Ortin:2015}
\begin{equation}
\sqrt{-g'} = \sqrt{-g}(1 + \tfrac{1}{2} h +\tfrac{1}{8}h^2 - \tfrac{1}{4}h_{\mu\nu}h^{\mu\nu} + \cdots),
\end{equation}
and Ricci scalar as $R' = R^{(0)}+R^{(1)}+R^{(2)}$, where
\begin{equation}
\begin{split}
R^{(0)} & = R, \\
R^{(1)} & = -h^{\mu\nu}R_{\mu\nu} + \nabla_\mu \nabla_\nu h^{\mu\nu}
 - \nabla_\mu \nabla^\mu h, \\
 R^{(2)} & = 2h^\mu{}_\rho h^{\rho \nu}R_{\mu\nu} - h^{\mu\nu}h^{\rho \lambda}R_{\mu\rho\nu\lambda} + h^{\mu\nu}\nabla_\mu \nabla_\nu h - \tfrac{1}{4}\nabla_\mu h \nabla^\mu h - \nabla_\mu h^{\mu\nu}\nabla_\rho h_\nu{}^\rho \\
& + \nabla_\mu h^{\mu\nu} \nabla_\nu h - 2h^{\mu\nu}\nabla_\rho \nabla_\mu h_{\nu}{}^\rho + h_{\mu\nu}\nabla_\rho \nabla^\rho h^{\mu\nu} - \tfrac{1}{2}\nabla_\mu h_{\nu\rho}\nabla^\rho h^{\mu\nu} +\tfrac{3}{4}\nabla_\rho h_{\mu\nu}\nabla^\rho h^{\mu\nu}.
\end{split}
\end{equation}
The Einstein–Hilbert Lagrangian can then be expressed as
\begin{equation}
\frac{1}{2\kappa_4^2}\frac{\sqrt{-g'}}{\sqrt{-g}}R' = R^{(0)} + [R^{(1)} + \tfrac{1}{2}h R^{(0)}] + [(\tfrac{1}{8}h^2 - \tfrac{1}{4}h_{\mu\nu}h^{\mu\nu})R^{(0)} + \tfrac{1}{2}hR^{(1)} + R^{(2)}].
\end{equation}

The leading order term is the background Einstein–Hilbert Lagrangian
\begin{equation}
\mathcal L_h^{(0)} = \frac{\sqrt{-g}}{2\kappa_4^2}R,
\end{equation}
while, by dropping total derivatives, the linear term can be written as 
\begin{equation}
\mathcal L_h^{(1)} = \frac{\sqrt{-g}}{2\kappa_4^2}\Big[\tfrac{1}{2}hR-h^{\mu\nu}R_{\mu\nu}\Big],
\end{equation}
Using
\begin{equation}
[\nabla_\mu, \nabla_\nu] h^\alpha{}_\beta = R^\alpha{}_{\rho \mu \nu}h^\rho{}_\beta - R^\rho{}_{\beta \mu\nu}h^\alpha{}_\rho,
\end{equation}
the $\sqrt{-g}R^{(2)}$ part of the second order term can be written up to total derivatives as
\begin{equation}
\sqrt{-g}R^{(2)} = \sqrt{-g}[h^\mu{}_\rho h^{\rho \nu}R_{\mu\nu} - \tfrac{1}{4}\nabla_\mu h \nabla^\mu h -\tfrac{1}{4}\nabla_\rho h_{\mu\nu}\nabla^\rho h^{\mu\nu} + \tfrac{1}{2} \nabla_\rho h^{\mu\nu}\nabla_\mu h_\nu{}^\rho].
\end{equation}
Combining this with the other second-order components gives
\begin{equation}
\begin{split}
\mathcal L_h^{(2)} & = \frac{\sqrt{-g}}{2\kappa_4^2}\bigg[\tfrac{1}{4}\nabla_\mu h \nabla^\mu h -\tfrac{1}{4}\nabla_\rho h_{\mu\nu}\nabla^\rho h^{\mu\nu} + \tfrac{1}{2}\nabla_\rho h^{\mu\nu}\nabla_\mu h_\nu{}^\rho -\tfrac{1}{2}\nabla_\mu h \nabla_\nu h^{\mu\nu} \\
& + \tfrac{1}{2} h^\mu{}_\rho h^{\rho \nu}R_{\mu\nu} +\tfrac{1}{2} h^{\mu\nu}h^{\rho \lambda}R_{\mu\rho\nu\lambda} -\tfrac{1}{2}hh^{\mu\nu}R_{\mu\nu} + \tfrac{1}{8}h^2R - \tfrac{1}{4}h_{\mu\nu}h^{\mu\nu}R\bigg].
\end{split}
\end{equation}

The kination field Lagrangian can be expanded in perturbations and written as $\mathcal L_\varphi = \mathcal L_\varphi^{(0)} + \mathcal L_\varphi^{(1)} + \mathcal L_\varphi^{(2)}$ where
\begin{equation}
\begin{split}
\mathcal L_\varphi^{(0)} & = -\frac{\sqrt{-g}}{2\kappa_4^2}R, \\
\mathcal L_\varphi^{(1)} & = -\frac{\sqrt{-g}}{2\kappa_4^2}\bigg[X^\mu \partial_\mu \varphi - h^{\mu\nu}R_{\mu\nu} + \tfrac{1}{2}hR\bigg], \\
\mathcal L^{(2)}_\varphi & = -\frac{\sqrt{-g}}{2\kappa_4^2}\bigg[\tfrac{1}{8}Rh^2 - \tfrac{1}{4}Rh_{\mu\nu}h^{\mu\nu} + h^\mu{}_\rho h^{\rho \nu}R_{\mu\nu} -\tfrac{1}{2}h h^{\mu\nu}R_{\mu\nu} \\
& \ \ \ \ \ \ \ \ \ \ \ \ \ \ + \tfrac{1}{2}\partial^\mu \varphi \partial_\mu \varphi - h^{\mu\nu}B_\mu \partial_\nu \varphi + \tfrac{1}{2}h B^\mu \partial_\mu \varphi\bigg].
\end{split}
\end{equation}

The total Lagrangian at each order in the perturbations is given by combining the contributions from the Einstein–Hilbert term and the scalar kinetic term $\mathcal L^{(i)} = \mathcal L^{(i)}_h + \mathcal L^{(i)}_\varphi$. The leading order term $\mathcal L^{(0)}$ vanishes since the background kination  solution is a solution to the Einstein equation. Meanwhile, most linear Lagrangian terms cancel with the remaining one being a total derivative
\begin{equation}
\mathcal L^{(1)} = -\frac{\sqrt{-g}}{2\kappa_4^2}B^\mu \partial_\mu \varphi = -\frac{\sqrt{-g}}{2\kappa_4^2} \nabla_\mu(B^\mu \varphi).
\end{equation}
The only non-vanishing terms are found in the bilinear Lagrangian
\begin{equation}\label{kination_pert}
\begin{split}
\mathcal L^{(2)} & =  \frac{\sqrt{-g}}{2\kappa_4^2}\bigg[\tfrac{1}{4}\nabla_\mu h \nabla^\mu h -\tfrac{1}{4}\nabla_\rho h_{\mu\nu}\nabla^\rho h^{\mu\nu} + \tfrac{1}{2}\nabla_\mu h^{\mu\nu}\nabla_\rho h_\nu{}^\rho -\tfrac{1}{2}\nabla_\mu h \nabla_\nu h^{\mu\nu} \\
& - \tfrac{1}{2}\partial^\mu \varphi \partial_\mu \varphi + h^{\mu\nu}B_\mu \partial_\nu \varphi - \tfrac{1}{2}h B^\mu \partial_\mu \varphi\bigg].
\end{split}
\end{equation}
This matches the Lagrangian for the metric perturbations and the dilaton acquired from the compactification of perturbations of the Kasner solution in Eq. \eqref{mainaction}.

\subsection*{Gauge Invariance of 4d Action}

 It is instructive to verify that the action is indeed gauge invariant. We do this by using the form of the transformations before the Weyl transformation was performed (i.e starting with Eq. \ref{drty}) and focusing for simplicity only on the variation that will be linear in the dilaton. The transformations act on the dilaton and scalar kinetic terms as
\begin{equation}
\begin{split}
\delta \mathcal L_{\phi\phi} & = \tfrac{8}{3}\zeta^3 \nabla^\mu \phi\nabla_\mu(\zeta^{-1}\xi_0) \\
& = \tfrac{16}{9}\zeta\xi_0 \nabla_0 \phi+\tfrac{8}{3}\zeta^2\nabla^\mu \phi\nabla_\mu \xi_0 \\
& = \tfrac{32}{27}\zeta^{-1}\phi\xi_0 +\tfrac{16}{9}\zeta \phi \nabla_0 \xi_0 - \tfrac{8}{3}\zeta^2 \phi \nabla_\mu \nabla^\mu \xi_0,
\end{split}
\end{equation}
where the first term on the second line is equivalent to $B^\mu\nabla_\mu \phi$. Meanwhile, the term in the action that we need to vary is
\begin{equation}
\begin{split}
\mathcal L_{h\phi} & = -\tfrac{4}{3}\zeta^2 \phi \nabla_0 h - \tfrac{8}{3}\zeta^2h_0^\mu \nabla_\mu \phi \\
& = \tfrac{4}{3}\zeta^2 h \nabla_0 \phi -\tfrac{32}{9}\zeta \phi h_{00}+\tfrac{8}{3}\zeta^2 \phi \nabla_\mu h_0^\mu.
\end{split}
\end{equation}
The variation has to be done carefully with
\begin{equation}
\delta h_{00} = B^\mu B^\nu \delta h_{\mu\nu} = 2 B^\mu B^\nu \nabla_\mu \xi_\nu = 2\nabla_0 \xi_0 + \tfrac{8}{3}\zeta^{-2}\xi_0,
\end{equation}
as well as 
\begin{equation}
\delta h = 2\nabla_\rho \xi^\rho - 8 \zeta^{-1}\xi_0.
\end{equation}
and additionally
\begin{equation}
\begin{split}
\delta (\nabla_\mu h_0^\mu) & = \nabla_\mu B^\alpha \nabla_\alpha \xi^\mu + \nabla_\mu  B^\alpha \nabla^\mu \xi_\alpha -2\nabla_\mu(B^\mu \zeta^{-1}\xi_0) \\
& = B^\alpha \nabla_\mu \nabla_\alpha \xi^\mu + \Gamma^\alpha_{0\mu}\nabla_\alpha \xi^\mu + \nabla_\mu \nabla^\mu \xi_0 + \tfrac{1}{3}\nabla_\mu(\zeta^{-2}\xi^\mu) +4\zeta^{-3}\xi_0-2\zeta^{-1}\nabla_0\xi_0 \\
& = B^\alpha \nabla_\mu \nabla_\alpha \xi^\mu + \nabla_\mu \nabla^\mu \xi_0 +\tfrac{40}{9}\zeta^{-3}\xi_0 - 2\zeta^{-1}\nabla_0 \xi_0.
\end{split}
\end{equation}
Using these we see that
\begin{equation}
\begin{split}
\delta \mathcal L_{h\phi} & = -\tfrac{8}{3}\zeta^2 \phi B^\alpha \nabla_\alpha \nabla_\rho \xi^\rho -\tfrac{64}{9}\zeta^{-1}\phi \xi_0 +\tfrac{32}{3}\zeta\phi \nabla_0 \xi_0 \\
& \ \ \ - \tfrac{64}{9}\zeta\phi \nabla_0 \xi_0 -\tfrac{256}{27}\zeta^{-1}\phi \xi_0 \\
& \ \ \ +\tfrac{8}{3}\zeta^2 \phi B^\alpha \nabla_\mu \nabla_\alpha \xi^\mu + \tfrac{8}{3}\zeta^2 \phi \nabla_\mu \nabla^\mu \xi_0 +\tfrac{320}{27}\zeta^{-1}\phi \xi_0 -\tfrac{16}{3}\zeta\phi \nabla_0\xi_0 \\
& = -\tfrac{8}{3}\zeta^2 \phi B^\alpha \nabla_\alpha \nabla_\rho \xi^\rho - \tfrac{64}{9}\zeta\phi \nabla_0 \xi_0 -\tfrac{128}{27}\zeta^{-1}\phi\xi_0+\tfrac{8}{3}\zeta^2 \phi B^\alpha \nabla_\mu \nabla_\alpha \xi^\mu + \tfrac{8}{3}\zeta^2 \phi \nabla_\mu \nabla^\mu \xi_0 \\
& = \tfrac{8}{3} \zeta^2 \phi R_{\rho 0}\xi^\rho -\tfrac{128}{27}\zeta^{-1}\phi \xi_0 - \tfrac{16}{9}\zeta \phi \nabla_0 \xi_0 + \tfrac{8}{3}\zeta^2 \phi \nabla_\mu \nabla^\mu \xi_0 \\
& = -\tfrac{32}{27}\zeta^{-1}\phi \xi_0 - \tfrac{16}{9}\zeta\phi \nabla_0 \xi_0 + \tfrac{8}{3}\zeta^2 \phi \nabla_\mu \nabla^\mu \xi_0,
\end{split}
\end{equation}
where we used that $R_{0\rho}\xi^\rho = \tfrac{4}{3}\zeta^{-3}\xi_0$. Now we see that $\delta \mathcal L_{h \phi} + \delta \mathcal L_{\phi \phi} = 0$ for the variation that is proportional to the dilaton. One can similarly show that the other variations also vanish. 

\bibliography{biblist}

@article{MUELLER199037,
title = {Rolling radii and a time-dependent dilation},
journal = {Nuclear Physics B},
volume = {337},
number = {1},
pages = {37-48},
year = {1990},
issn = {0550-3213},
doi = {https://doi.org/10.1016/0550-3213(90)90249-D},
url = {https://www.sciencedirect.com/science/article/pii/055032139090249D},
author = {Mark Mueller},
}

@article{Sugiyama:2020mbu,
    author = "Sugiyama, Yuuki and Yamamoto, Kazuhiro and Kobayashi, Tsutomu",
    title = "{Gravitational waves in Kasner spacetimes and Rindler wedges in Regge-Wheeler gauge: Formulation of Unruh effect}",
    eprint = "2012.15004",
    archivePrefix = "arXiv",
    primaryClass = "gr-qc",
    reportNumber = "RUP-20-35",
    doi = "10.1103/PhysRevD.103.083503",
    journal = "Phys. Rev. D",
    volume = "103",
    number = "8",
    pages = "083503",
    year = "2021"
}

@article{Grimm:2004ua,
    author = "Grimm, Thomas W. and Louis, Jan",
    title = "{The Effective action of type IIA Calabi-Yau orientifolds}",
    eprint = "hep-th/0412277",
    archivePrefix = "arXiv",
    doi = "10.1016/j.nuclphysb.2005.04.007",
    journal = "Nucl. Phys. B",
    volume = "718",
    pages = "153--202",
    year = "2005"
}

@article{Balasubramanian:2005zx,
    author = "Balasubramanian, Vijay and Berglund, Per and Conlon, Joseph P. and Quevedo, Fernando",
    title = "{Systematics of moduli stabilisation in Calabi-Yau flux compactifications}",
    eprint = "hep-th/0502058",
    archivePrefix = "arXiv",
    reportNumber = "DAMTP-2005-10, UNH-05-01, UPR-1109-T",
    doi = "10.1088/1126-6708/2005/03/007",
    journal = "JHEP",
    volume = "03",
    pages = "007",
    year = "2005"
}

@article{Kachru:2003aw,
    author = "Kachru, Shamit and Kallosh, Renata and Linde, Andrei D. and Trivedi, Sandip P.",
    title = "{De Sitter vacua in string theory}",
    eprint = "hep-th/0301240",
    archivePrefix = "arXiv",
    reportNumber = "SLAC-PUB-9630, SU-ITP-03-01, TIFR-TH-03-03",
    doi = "10.1103/PhysRevD.68.046005",
    journal = "Phys. Rev. D",
    volume = "68",
    pages = "046005",
    year = "2003"
}

@article{Apers:2024ffe,
    author = "Apers, Fien and Conlon, Joseph P. and Copeland, Edmund J. and Mosny, Martin and Revello, Filippo",
    title = "{String Theory and the First Half of the Universe}",
    eprint = "2401.04064",
    archivePrefix = "arXiv",
    primaryClass = "hep-th",
    month = "1",
    year = "2024"
}

@article{Joyce:1996cp,
    author = "Joyce, Michael",
    title = "{Electroweak Baryogenesis and the Expansion Rate of the Universe}",
    eprint = "hep-ph/9606223",
    archivePrefix = "arXiv",
    reportNumber = "CERN-TH-96-098, CERN-TH-96-98",
    doi = "10.1103/PhysRevD.55.1875",
    journal = "Phys. Rev. D",
    volume = "55",
    pages = "1875--1878",
    year = "1997"
}

@article{Cicoli:2023opf,
    author = "Cicoli, Michele and Conlon, Joseph P. and Maharana, Anshuman and Parameswaran, Susha and Quevedo, Fernando and Zavala, Ivonne",
    title = "{String cosmology: From the early universe to today}",
    eprint = "2303.04819",
    archivePrefix = "arXiv",
    primaryClass = "hep-th",
    doi = "10.1016/j.physrep.2024.01.002",
    journal = "Phys. Rept.",
    volume = "1059",
    pages = "1--155",
    year = "2024"
}

@article{Giddings:2001yu,
    author = "Giddings, Steven B. and Kachru, Shamit and Polchinski, Joseph",
    title = "{Hierarchies from fluxes in string compactifications}",
    eprint = "hep-th/0105097",
    archivePrefix = "arXiv",
    reportNumber = "SLAC-PUB-8807, NSF-ITP-01-37, SU-ITP-01-16",
    doi = "10.1103/PhysRevD.66.106006",
    journal = "Phys. Rev. D",
    volume = "66",
    pages = "106006",
    year = "2002"
}

@article{Revello:2023hro,
    author = "Revello, Filippo",
    title = "{Attractive (s)axions: cosmological trackers at the boundary of moduli space}",
    eprint = "2311.12429",
    archivePrefix = "arXiv",
    primaryClass = "hep-th",
    doi = "10.1007/JHEP05(2024)037",
    journal = "JHEP",
    volume = "05",
    pages = "037",
    year = "2024"
}

@article{Seo:2024qzf,
    author = "Seo, Min-Seok",
    title = "{Asymptotic behavior of saxion-axion system in stringy quintessence model}",
    eprint = "2403.07307",
    archivePrefix = "arXiv",
    primaryClass = "hep-th",
    month = "3",
    year = "2024"
}

@article{Conlon:2022pnx,
    author = "Conlon, Joseph P. and Revello, Filippo",
    title = "{Catch-me-if-you-can: the overshoot problem and the weak/inflation hierarchy}",
    eprint = "2207.00567",
    archivePrefix = "arXiv",
    primaryClass = "hep-th",
    doi = "10.1007/JHEP11(2022)155",
    journal = "JHEP",
    volume = "11",
    pages = "155",
    year = "2022"
}

@article{Shiu:2023nph,
    author = "Shiu, Gary and Tonioni, Flavio and Tran, Hung V.",
    title = "{Accelerating universe at the end of time}",
    eprint = "2303.03418",
    archivePrefix = "arXiv",
    primaryClass = "hep-th",
    doi = "10.1103/PhysRevD.108.063527",
    journal = "Phys. Rev. D",
    volume = "108",
    number = "6",
    pages = "063527",
    year = "2023"
}

@article{Shiu:2023fhb,
    author = "Shiu, Gary and Tonioni, Flavio and Tran, Hung V.",
    title = "{Late-time attractors and cosmic acceleration}",
    eprint = "2306.07327",
    archivePrefix = "arXiv",
    primaryClass = "hep-th",
    doi = "10.1103/PhysRevD.108.063528",
    journal = "Phys. Rev. D",
    volume = "108",
    number = "6",
    pages = "063528",
    year = "2023"
}

@article{Baumann:2009ni,
    author = "Baumann, Daniel and McAllister, Liam",
    title = "{Advances in Inflation in String Theory}",
    eprint = "0901.0265",
    archivePrefix = "arXiv",
    primaryClass = "hep-th",
    doi = "10.1146/annurev.nucl.010909.083524",
    journal = "Ann. Rev. Nucl. Part. Sci.",
    volume = "59",
    pages = "67--94",
    year = "2009"
}

@article{Brandenberger:2023ver,
    author = "Brandenberger, Robert",
    title = "{Superstring cosmology \textemdash{} a complementary review}",
    eprint = "2306.12458",
    archivePrefix = "arXiv",
    primaryClass = "hep-th",
    doi = "10.1088/1475-7516/2023/11/019",
    journal = "JCAP",
    volume = "11",
    pages = "019",
    year = "2023"
}

@article{Apers:2022cyl,
    author = "Apers, Fien and Conlon, Joseph P. and Mosny, Martin and Revello, Filippo",
    title = "{Kination, meet Kasner: on the asymptotic cosmology of string compactifications}",
    eprint = "2212.10293",
    archivePrefix = "arXiv",
    primaryClass = "hep-th",
    doi = "10.1007/JHEP08(2023)156",
    journal = "JHEP",
    volume = "08",
    pages = "156",
    year = "2023"
}

@article{Fierz:1939,
    author = "Fierz, Markus and Pauli,Wolfgang",
    title = "{Relativistic Wave Equations for Particles of Arbitrary Spin in an Electromagnetic Field}",
    eprint = "",
    archivePrefix = "",
    primaryClass = "",
    reportNumber = "",
    doi = "10.1098/rspa.1939.0140",
    journal = "Proc. R. Soc. Lond. A",
    volume = "173",
    pages = "211",
    year = "1939"
}

@book{Ortin:2015,
    author = "Ort\'in, Tom\'as",
    title = "{Gravity and Strings}",
    publisher = "Cambridge University Press",
    chapter = "3",
    isbn = "978-0521768139",
    pages = "133",
    year = "2015"
}

@article{Han:1998sg,
    author = "Han, Tao and Lykken, Joseph D. and Zhang, Ren-Jie",
    title = "{On Kaluza-Klein states from large extra dimensions}",
    eprint = "hep-ph/9811350",
    archivePrefix = "arXiv",
    reportNumber = "MADPH-98-1092, FERMILAB-PUB-98-364",
    doi = "10.1103/PhysRevD.59.105006",
    journal = "Phys. Rev. D",
    volume = "59",
    pages = "105006",
    year = "1999"
}

\end{document}